\documentclass[pre,aps,twocolumn,showpacs,floatfix,10pt,reprint]{revtex4-1}
\usepackage[dvips]{graphicx}
\usepackage{amsmath}
\usepackage{amssymb}
\usepackage[nice]{nicefrac}
\usepackage{color}
\usepackage[hidelinks]{hyperref}
\hypersetup{colorlinks=false}

\usepackage{times}

\begin{document}

\title{Underdamped stochastic harmonic oscillator}

\author{Bart{\l}omiej Dybiec}
\email{bartek@th.if.uj.edu.pl}
\affiliation{Marian Smoluchowski Institute of Physics, and Mark Kac Center for Complex Systems Research, Jagiellonian University, ul. St. {\L}ojasiewicza 11, 30--348 Krak\'ow, Poland}

\author{Igor M. Sokolov}
\email{igor.sokolov@physik.hu-berlin.de}
\affiliation{Institut f\"ur Physik, Humboldt-Universit\"at zu Berlin, Newtonstrasse 15, D--12489 Berlin, Germany}

\author{Ewa Gudowska-Nowak}
\email{gudowska@th.if.uj.edu.pl}
\affiliation{Marian Smoluchowski Institute of Physics, and Mark Kac Center for Complex Systems Research, Jagiellonian University, ul. St. {\L}ojasiewicza 11, 30--348 Krak\'ow, Poland}

\date{\today}
\begin{abstract}
We investigate the distribution of potential and kinetic energy in stationary states of the linearly damped stochastic oscillator driven by L\'evy noises.
In the long time limit distributions of kinetic and potential energies of the oscillator follow the power-law asymptotics and do not fulfill the equipartition theorem. The partition of the mechanical energy is controlled by the damping coefficient.
In the limit of vanishing damping a stochastic analogue of the equipartition theorem can be proposed, namely the statistical properties of potential and kinetic energies attain distributions characterized by the same widths.
For larger damping coefficient the larger fraction of energy is stored in its potential form. In the limit of very strong damping the contribution of kinetic energy becomes negligible.
Finally, we demonstrate that the ratio of instantaneous kinetic and potential energies, which signifies departure from the mechanical energy equipartition, follows universal power-law asymptotics, regardless of the symmetric $\alpha$-stable noise parameters.
Altogether our investigations clearly indicate strongly non-equilibrium character of L\'evy-stable fluctuations with the stability index $\alpha < 2$.

\end{abstract}

\maketitle

\section{Introduction\label{sec:introduction}}

A damped harmonic oscillator under influence of noise is one of the fundamental conceptual models in non-equilibrium statistical physics \cite{doob1942,west1982,chechkin2000linear,gitterman2005noisy,lin2011undamped}, broadly used to describe relaxation phenomena in the linear regime.
Displacement $x(t)$ from the minimum of the potential $V(x)$ is described by the Langevin equation
\begin{equation}
m\frac{d^2x(t)}{dt^2}=-\gamma \frac{dx(t)}{dt} - \lambda x(t) + \sqrt{2 C }\xi(t),
\label{eq:newton}
\end{equation}
in which the interaction with the environment is separated into deterministic dissipative force $\gamma \frac{dx(t)}{dt}$, describing damping, and a noise term $\xi(t)$ describing all the complexity of the interaction between the test particle (or mode) with the rest of the system.
In the situation of contact of the system with a single heat bath the (intrinsic) noise corresponding to linear (Stokes) friction has to be assumed Gaussian and white, i.e. it fulfills $\langle \xi(t) \rangle=0$ and $\langle\xi(t) \xi(s) \rangle=\delta(t-s)$.
The coefficient $C$ in Eq.~(\ref{eq:newton}) is given by $C=\gamma \frac{k_B T}{m}$,
where $m$ stands for the particle's (effective) mass, $T$ is the system's temperature, and $k_B$ is the Boltzmann constant.
The joint probability density $P=P(x,v,t)$ evolves  according to the Kramers equation \cite{risken1984}
\small
\begin{equation}
 \frac{\partial P }{\partial t}=\left[- v \frac{\partial}{\partial x} +\frac{\partial}{\partial v}\left( \gamma v + \frac{V'(x)}{m} \right) +\gamma \frac{k_B T}{m}\frac{\partial^2}{\partial v^2} \right]P.
 \label{eq:kk}
\end{equation}
\normalsize
The stationary solution of Eq.~(\ref{eq:kk}) has the canonical Boltzmann-Gibbs form \cite{chandrasekhar1943,risken1984}
\begin{equation}
 P(x,v) = N \exp\left[ - \frac{1}{k_B T} \left( \frac{m v^2}{2} + {V(x)} \right) \right],
 \label{eq:st}
\end{equation}
and factorizes, making position and velocity to be statistically independent random variables.
If $V(x)=\lambda \frac{x^2}{2}$ as in Eq.~(\ref{eq:newton}), the stationary solution is a 2D, elliptically contoured normal density.  Moreover, the average energy of the oscillator in  its stationary (equilibrium) state is then given by the classical equipartition value $\langle \mathcal{E} \rangle=k_BT$ with $\mathcal{E}_k=mv^2/2$ and $\mathcal{E}_p=\lambda x^2/2$
\begin{equation}
 \left\langle \frac{m v^2}{2} \right\rangle = \left\langle \frac{\lambda x^2}{2} \right\rangle = \frac{k_B T}{2}.
 \label{eq:equipartition}
\end{equation}
Within stochastic approach, the equipartition theorem originates due to properties of the noise.

From a physical point of view, the ``whiteness'' of the noise is a
consequence of the large number of statistically independent interactions of a test particle with molecules of heat bath which are
bounded in time. Its Gaussian character arises due to the assumption that the
interactions are bounded in their strength.  In many far-from-
equilibrium situations the second assumption fails. The noise still
can be considered as white but is now described by
heavy-tailed distributions, often of the $\alpha$-stable L\'evy
type \cite{janicki1994,*chechkin2006,*dubkov2008}.  Heavy-tailed
fluctuations of the $\alpha$-stable type have been observed in turbulent fluid flows
\cite{shlesinger1993,*klafter1996,*solomon1993,
*delcastillonegrete1998}, magnetized plasmas \cite{chechkin2002b,
*delcastillonegrete2005}, optical lattices \cite{Katori1997},
heartbeat dynamics \cite{peng1993}, neural networks \cite{segev2002},
search on a folding polymers \cite{lomholt2005}, animal movement
\cite{Viswanathan1996}, climate dynamics \cite{ditlevsen1999b},
financial time series \cite{mantegna2000}, and even in spreading of
diseases and dispersal of banknotes \cite{brockmann2006}.
Such large fluctuations, not appearing in the description of equilibrium bath, can be attributed to the external forcing.
The white L\'evy noise which naturally appears in description of systems far-from-equilibrium
breaks the microscopic reversibility, and changes considerably the properties of the stationary states of the
system compared to equilibrium cases \cite{kusmierz2016breaking}.

Within the current work we assume that white Gaussian noise in Eq.~(\ref{eq:newton}) is replaced by external white L\'evy noise \cite{samorodnitsky1994,janicki1994,janicki1996,chechkin2006,dubkov2008,vahabi2013} but the dissipation is still of the Stokes type.
The behavior of a damped harmonic oscillator under the influence of noise with $\alpha<2$ is very different from those for the Gaussian case $\alpha=2$. Stationary densities are given by bivariate $\alpha$-stable densities \cite{press1972,samorodnitsky1994} for which lines of constant probability are not ellipses. Moreover, in the stationary state, velocity $v$ and position $x$ are not statistically independent \cite{samorodnitsky1994}. Stationary states for a particle moving in the parabolic potential driven by the white L\'evy noise reflect symmetries of the noise, i.e. they are given by the $\alpha$-stable densities both for symmetric \cite{jespersen1999,chechkin2002} and asymmetric noises \cite{dybiec2007d}. The same effect is observed for a 2D parabolic potential perturbed by the bi-variate L\'evy noise \cite{szczepaniec2014}.
In what follows we investigate distributions of kinetic and potential energies of a damped harmonic oscillator and distributions of ratio of their instantaneous values for the case of symmetric L\'evy noises.

\section{Model and results\label{sec:model}}

We examine the distributions of the kinetic and potential energy and of their ratio for the case of a damped harmonic oscillator driven by a white $\alpha$-stable noise $\zeta_\alpha(t)$:
\begin{equation}
m\ddot{x}(t)=-\gamma \dot{x}(t) - \lambda x(t) + \zeta_\alpha(t).
\label{HaO}
\end{equation}
The white $\alpha$-stable noise $\zeta_\alpha(t)$, which is a formal time derivative of the $\alpha$-stable motion $L_\alpha(t)$ \cite{janicki1994b}, results in stochastic increments which are distributed according to the symmetric $\alpha$-stable density whose characteristic function is given by \cite{samorodnitsky1994,janicki1994}
\begin{equation}
 \phi(k)=\exp\left[ -\sigma^\alpha |k|^\alpha \right].
 \label{eq:char}
\end{equation}
The parameter $\alpha$ ($0 < \alpha \leqslant 2$) is the so called stability index describing asymptotics of $\alpha$-stable densities which for $\alpha<2$ is of the power-law type
$p(x) \propto |x|^{-(\alpha+1)}$.
In the $\alpha=2$ limit the $\alpha$-stable noise is equivalent to the Gaussian white noise.
The strength of fluctuations in Eq.~(\ref{HaO}) is controlled by the scale parameter $\sigma$, see Eq.~(\ref{eq:char}) and its similarity properties are governed by the parameter $\alpha$.
Contrary to the $\alpha=2$ case, due to the divergence of the second moment $\langle v^2 \rangle$, for $0<\alpha<2$, there is no fluctuation-dissipation relation of the Smoluchowski-Sutherland-Einstein type \cite{smoluchowski1906,einstein1905,touchette2009,chechkin2009,gudowska2014}. Consequently, damping coefficient $\gamma$ and fluctuation intensity $\sigma$ can be viewed as independent parameters.
The parameters $m$ and $\lambda$ of free, undamped oscillator define the most convenient units in which the system can be described.
We choose $t_0 = \omega_0^{-1} = \sqrt{m/\lambda}$ to define the unit of time.
In the dimensionless time $t/t_0$ (for a brief explanation of units, cf. Appendix~\ref{sec:app-units}) the equation (\ref{HaO}) takes the form
\begin{equation}
\ddot{x}(t)=-\gamma \dot{x}(t) - x(t) + \sigma \zeta_\alpha(t),
\label{eq:dimless}
\end{equation}
with a damping $\gamma$ replacing the original frequency of dissipation $\gamma=\tilde{\gamma} t_0/m$ (for the clarity, we omit the tilde sign over the original constants). Here the prefactor $\sigma$ of $\zeta_\alpha$, measuring intensity of the noise is $\sigma=\tilde{\sigma} t_0^{1+1/\alpha}/m$. Moreover, in Eq.~(\ref{eq:dimless}), the white L\'evy noise $\zeta_\alpha(t)$ with the scale parameter set to unity is used.
The instantaneous kinetic and potential energies of the system are denoted by $\mathcal{E}_k = v^2/2$ and $\mathcal{E}_p = x^2/2$, respectively.

\begin{figure}[!ht]
\includegraphics[width=\columnwidth]{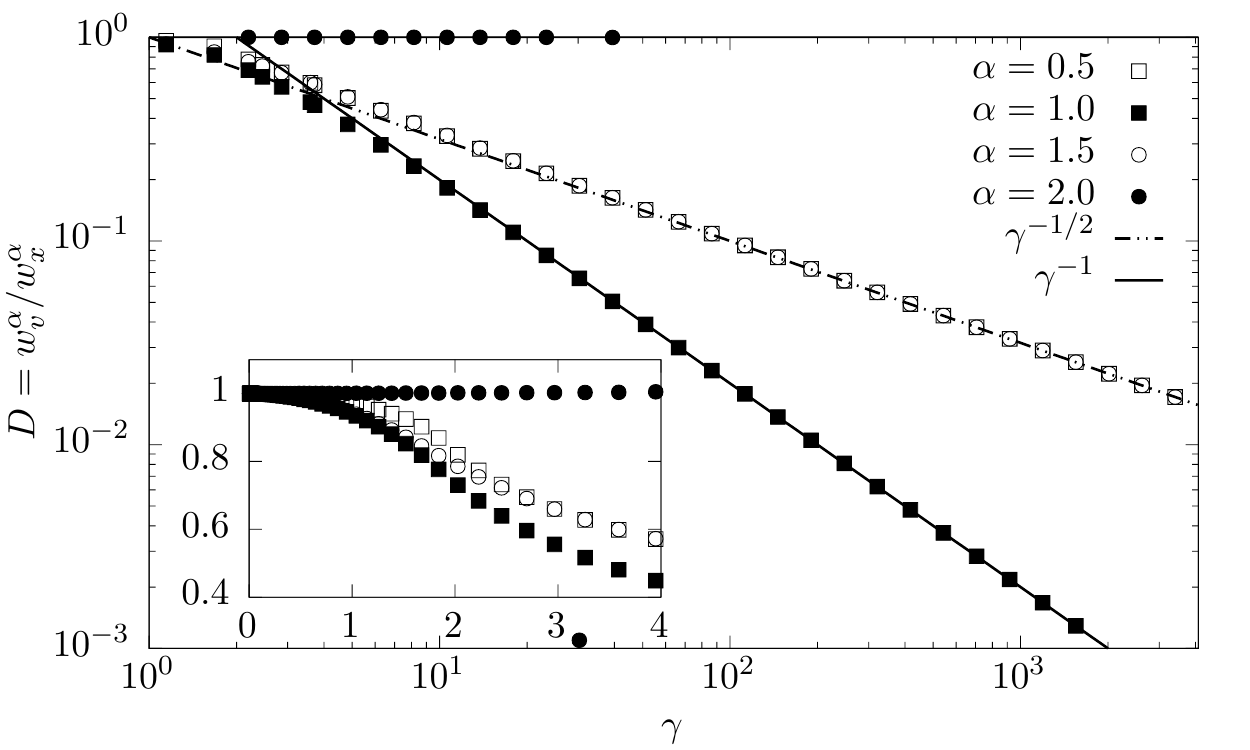}
 \caption{The quotient of the corresponding prefactors $D=\frac{w_v^\alpha}{w_x^\alpha}$ for various values of the stability index $\alpha$. Please note the double logarithmic scale in the main plot and the linear scale in the inset.}
 \label{fig:width}
\end{figure}

 The formal solution of
Eq.~(\ref{eq:dimless}) is
\begin{equation}
x(t) = F(t) + \int_{-\infty}^t G(t-t') \zeta_\alpha(t') dt',
\label{ForSol}
\end{equation}
where $G(t)$ is the Green's (response) function of the corresponding
process, and $F(t)$ is a decaying function (a solution of the
homogeneous equation under given initial conditions). The solution for
$v$ is given by
\begin{equation}
v(t) = F_v(t) + \int_{-\infty}^t G_v(t-t') \zeta_\alpha(t') dt',
\label{SolVel}
\end{equation}
where $G_v(t)$ is the Green's function of the velocity process
\begin{equation}
G_v(t)=\frac{d}{dt} G(t).
\end{equation}
In a stationary situation, $t \to \infty$, the $F$-functions in
Eqs.~(\ref{ForSol}) and (\ref{SolVel}) vanish. The Green's function of
Eq.~(\ref{HaO}) can be easily found e.g. via the Laplace
representation, and reads:
\begin{equation}
G(t)=\frac{\exp(-\gamma t/2)}{\sqrt{\omega_0^2-\gamma^2/4}} \sin\left[\sqrt{\omega_0^2-\gamma^2/4}\; t\right]
\end{equation}
for $\omega_0 =\sqrt{\lambda/m}=1 > \gamma/2$ (underdamped case),
\begin{equation}
G(t)= t \exp(-\gamma t/2)
\end{equation}
for $\omega_0 =1= \gamma/2$ (critical case)
and
\begin{equation}
G(t) = \frac{\exp(-\gamma t/2)}{\sqrt{\gamma^2/4-\omega_0^2}} \sinh \left[\sqrt{\gamma^2/4-\omega_0^2}\;t\right]
\end{equation}
for $\omega_0 =\sqrt{\lambda/m}=1 < \gamma/2$ (overdamped case).
Note that the functions $G(t)$ vanish both for $t=0$ and for
$t \to \infty$
so that
\begin{eqnarray}
\int_0^\infty G(t) \left[ \frac{d}{dt} G(t) \right] dt & = & \frac{1}{2}\int_0^\infty \left[ \frac{d}{dt} G^2(t) \right] dt \\
& = & \frac{1}{2} \left. G^2(t) \right|_0^\infty =0, \nonumber
\end{eqnarray}
i.e. $G(t)$ and $G_v(t)$ are orthogonal on $[0,\infty)$.

The characteristic function of the stationary distribution of $x$ and $v= \dot{x}$ is given by Eq.~(17) of Ref.~\cite{sokolov2010}:
\begin{equation}
f(k,q)=\exp\left[- \sigma^\alpha \int_0^\infty \left| kG(t) + qG_v(t)\right|^\alpha \right]
\label{JCF}
\end{equation}
where $G(t)$ is the Green's function for the homogeneous part of the equation of motion, see Eqs.~(6) -- (8) of Ref.~\cite{sokolov2010}, $G_v(t)=\frac{d} {dt} G(t)$.
Moreover, the Eq.~(\ref{JCF}) is the characteristic function of the 2D $\alpha$-stable density \cite{samorodnitsky1994}.

The marginal distributions of $x$ and $v$ have the characteristic functions $f_x(k)=f(k,0)$ and $f_v(q)=f(0,q)$ and are the L\'evy
stable ones with index $\alpha$ and scale parameters (widths)
\begin{equation}
w_x^\alpha = \sigma^\alpha \int_0^\infty |G(t)|^\alpha dt
\end{equation}
and
\begin{equation}
w_v^\alpha = \sigma^\alpha \int_0^\infty |G_v(t)|^\alpha dt.
\end{equation}
The corresponding integrals can be easily evaluated numerically for small and moderate values of $\gamma$ for any $\alpha >0$. Their asymptotic behavior for $\gamma \to 0$ and for $\gamma \to \infty$ will be discussed in the next subsection.

The large $|x|$ and $|v|$ asymptotics of the corresponding PDFs are
\begin{equation}
p_x(x) \propto \frac{w_x^\alpha}{|x|^{1+\alpha}}
\end{equation}
and
\begin{equation}
p_v(v) \propto \frac{w_v^\alpha}{|v|^{1+\alpha}},
\end{equation}
i.e. they are of the of the $\alpha$-stable type with the same stability index like the driving noise.

\subsection{Ratio of distribution widths}

The numerically calculated quotient of the corresponding prefactors
\begin{equation}
D=\frac{w_v^\alpha}{w_x^\alpha}
\label{eq:quotient}
\end{equation}
is depicted in the Fig.~\ref{fig:width} as a function of the damping coefficient $\gamma$.
Various curves correspond to different values of the stability index $\alpha$.
The inset shows small $\gamma$ dependence.

As derived in the Appendix~\ref{sec:app-ratio-width}, the ratio of distributions widths scales as
\begin{equation}
 D= \frac{w_v^\alpha}{w_x^\alpha}=
 \left\{
 \begin{array}{lcl}
 \gamma^{-\alpha} & \mbox{for} & 0< \alpha <1 \\
 \gamma^{\alpha-2} & \mbox{for} & 1 \leqslant \alpha \leqslant 2 \\
 \end{array}
 \right..
 \label{eq:dscaling}
\end{equation}
The formula~(\ref{eq:dscaling}), accounts also for $\alpha=2$ when  $ D=1$.

\begin{figure}[!ht]
\includegraphics[width=\columnwidth]{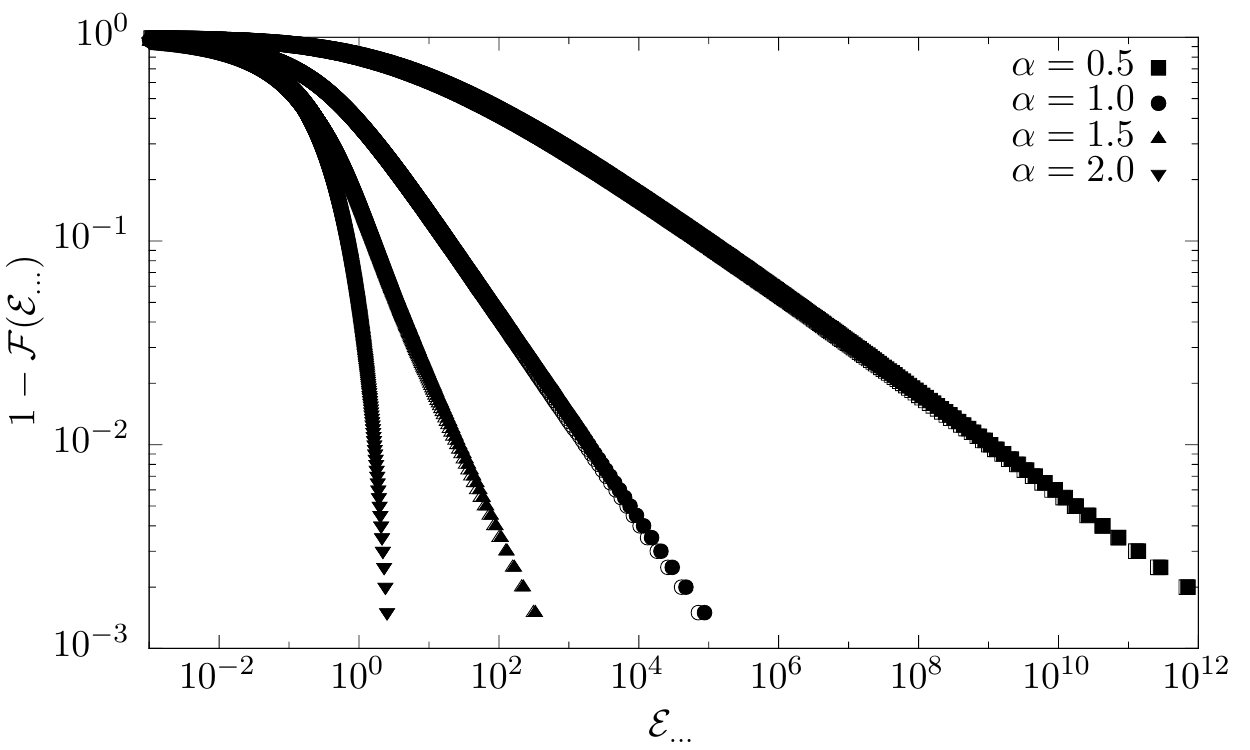} \\
\includegraphics[width=\columnwidth]{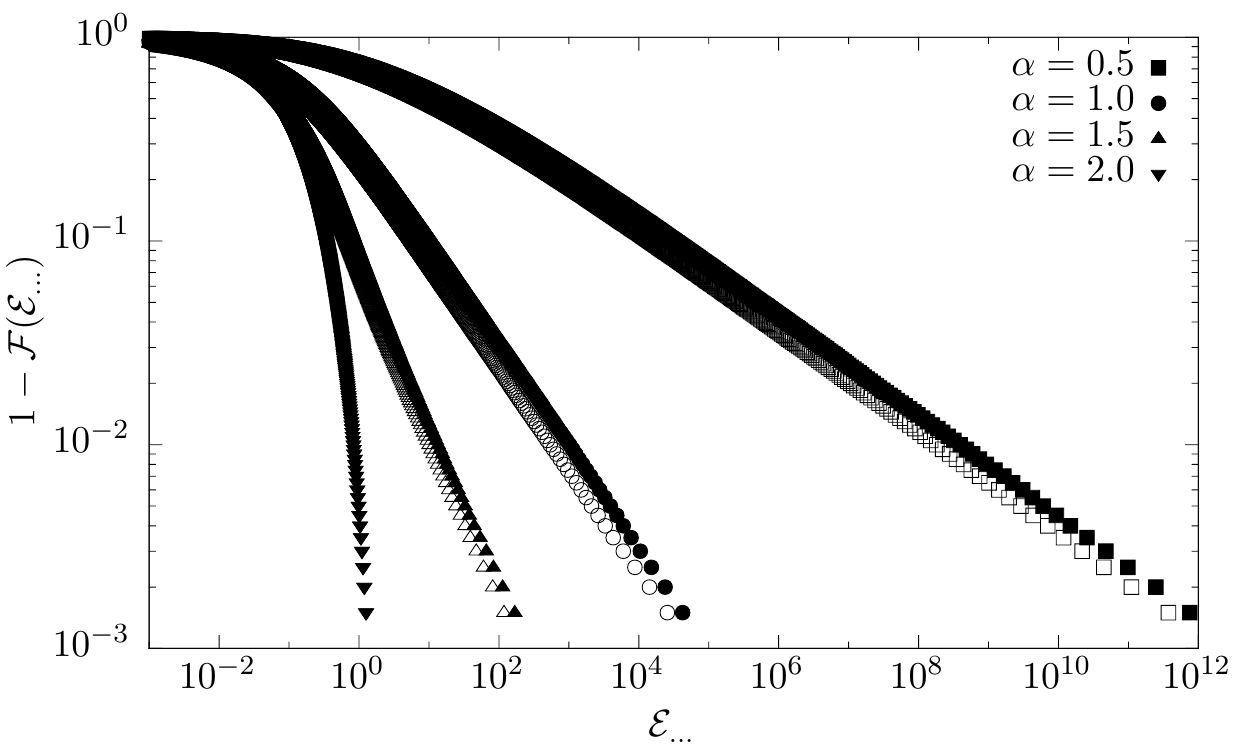} \\
\includegraphics[width=\columnwidth]{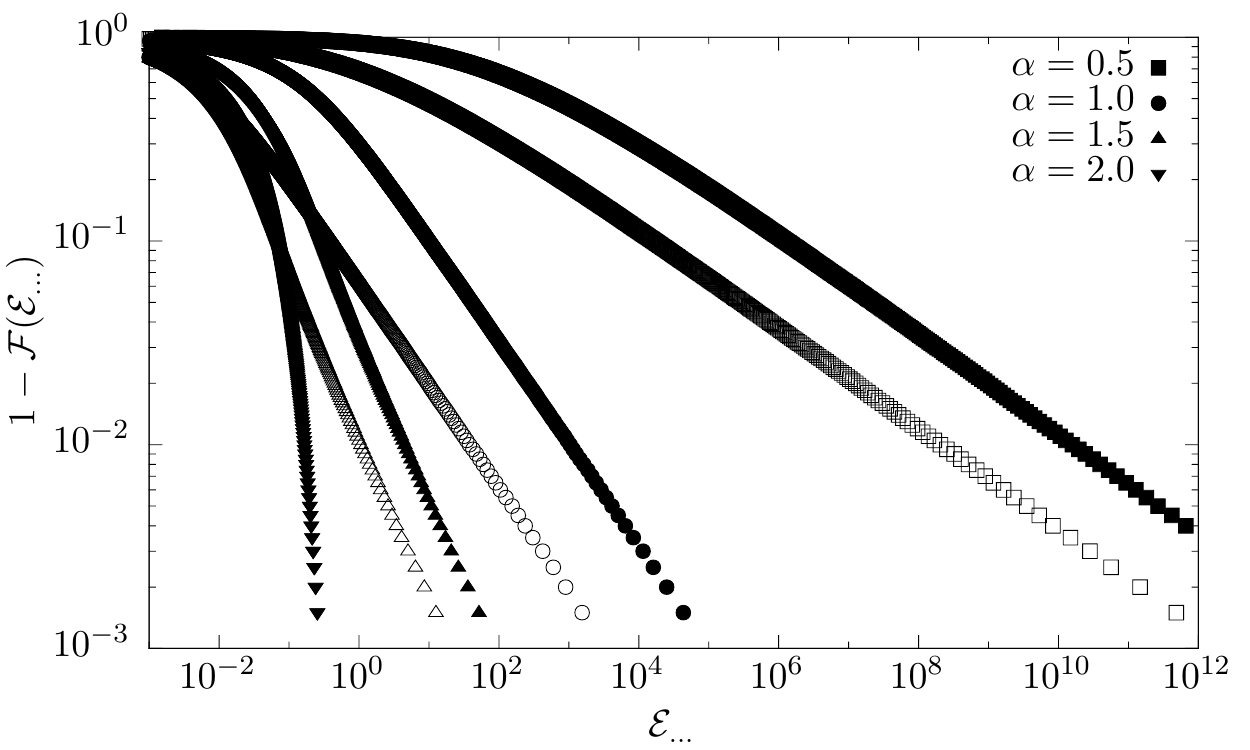} \\
 \caption{Distributions of the potential $\mathcal{E}_p$ (full symbols) and kinetic $\mathcal{E}_k$ (empty symbols) energies for $\gamma=1,2,10$ (from top to bottom). Various curves correspond to various values of the stability index $\alpha$.}
 \label{fig:ep-ek}
\end{figure}

Numerical simulations presented in Fig.~\ref{fig:width} perfectly confirm the scaling predicted by Eq.~(\ref{eq:dscaling}).
Please note, that results for $\alpha=0.5$ (empty squares) and $\alpha=1.5$ (empty circles) coincide.

\begin{figure}[!ht]
\includegraphics[width=\columnwidth]{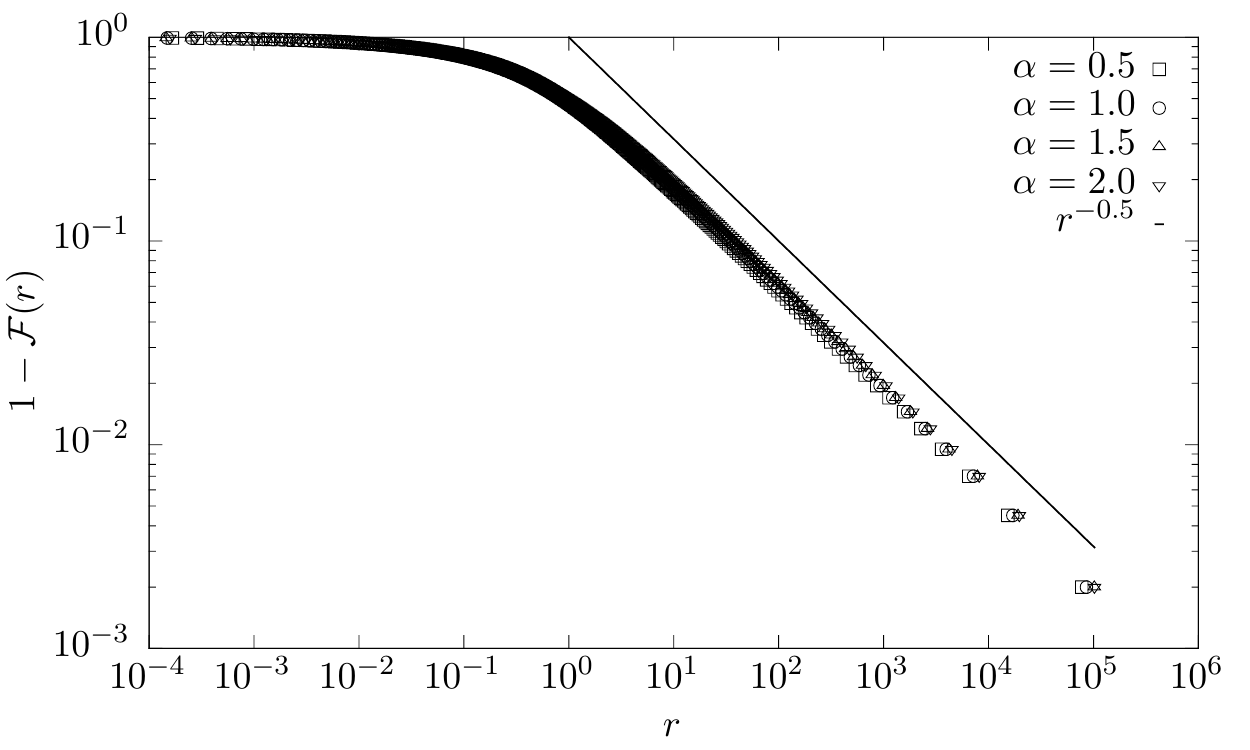} \\
\includegraphics[width=\columnwidth]{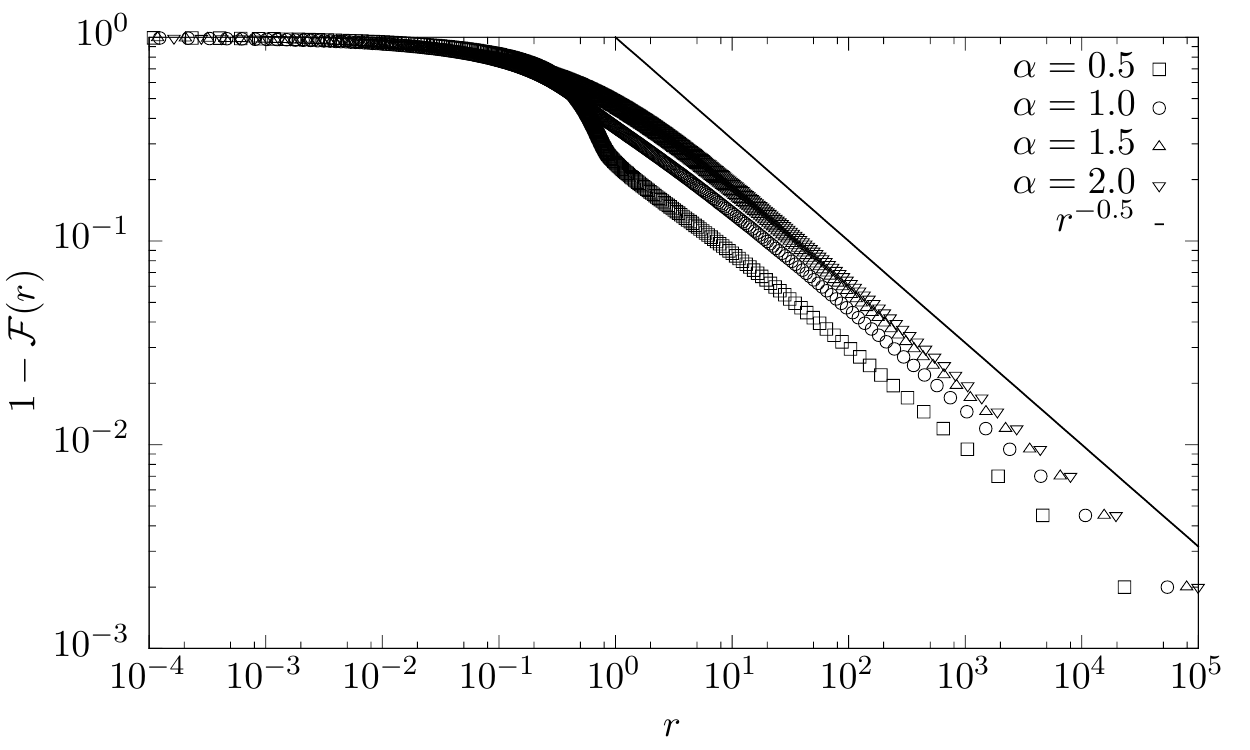} \\
\includegraphics[width=\columnwidth]{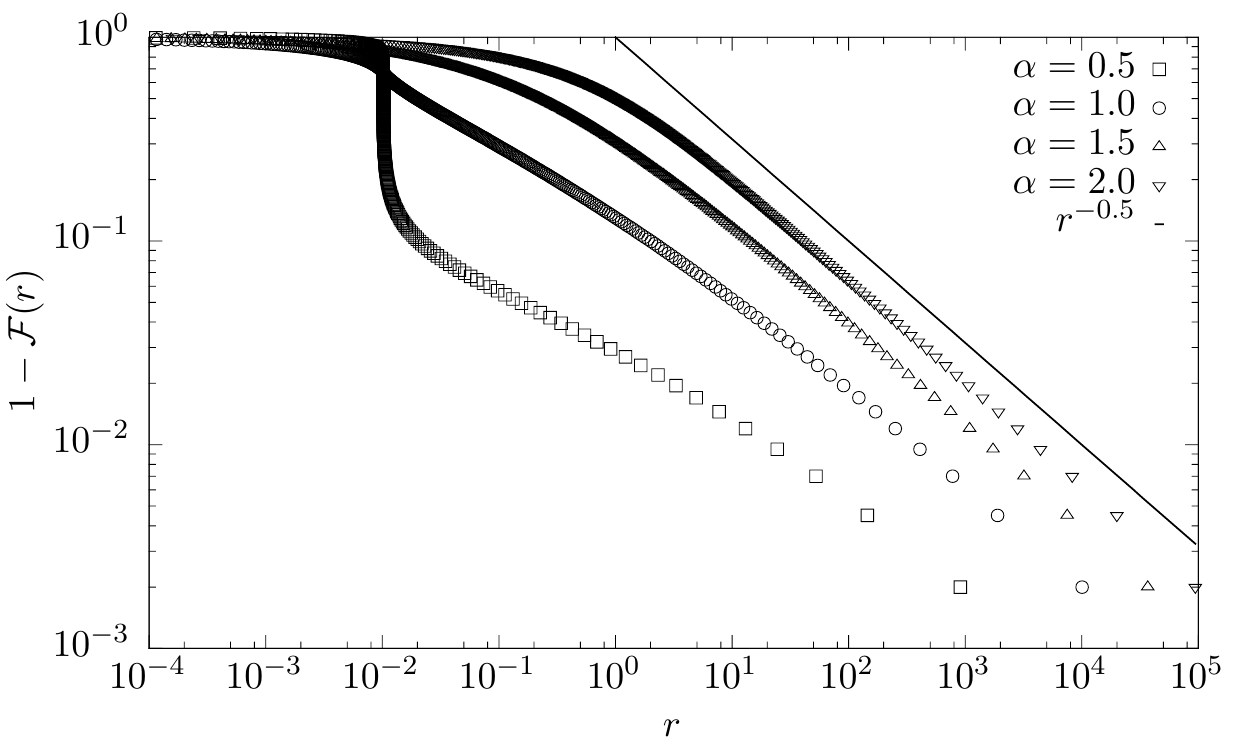} \\
 \caption{Distribution of the energy ratio $r=\mathcal{E}_k/\mathcal{E}_p$ for $\gamma=1,2,10$ (from top to bottom).}
 \label{fig:ratio}
\end{figure}

\subsection{Energy distributions}
Let us consider the distributions of the kinetic and the potential energies. Through the change of variables $x =\pm \sqrt{2\mathcal{E}_p}$, $v=\pm\sqrt{2\mathcal{E}_k}$
we get
\begin{equation}
p(\mathcal{E}_p) = p_x(\sqrt{2\mathcal{E}_p})\mathcal{E}_p^{-1/2} \propto \frac{w_x^\alpha}{\mathcal{E}_p^{1+\frac{\alpha}{2}}},
\end{equation}
\begin{equation}
p(\mathcal{E}_k) = p_v(\sqrt{2\mathcal{E}_k})\mathcal{E}_k^{-1/2} \propto \frac{w_v^\alpha}{\mathcal{E}_k^{1+\frac{\alpha}{2}}}.
\end{equation}
The total energy $\mathcal{E}$ in our units is proportional to the square of the amplitude of the phase space vector
$A=\sqrt{x^2+v^2}$: $E = A^2/2$.
Since this amplitude has a distribution
\begin{equation}
p(A) \propto \frac{1}{A^{1+\alpha}},
\end{equation}
the total energy has the same asymptotics as the kinetic or the potential one, and it is dominated by the potential energy in
the case of large $\gamma$.

Figure~\ref{fig:ep-ek} presents sample potential $\mathcal{E}_p$ (full symbols) and kinetic $\mathcal{E}_k$ energy (empty symbols) distributions for various values of the damping coefficient $\gamma$ ($\gamma=1,2,10$) and stability index $\alpha$ ($\alpha=0.5,1.0,1.5,2.0$).

It is interesting to discuss the behavior of the instantaneous quotient of the kinetic and potential energy
\begin{equation}
r = \frac{\mathcal{E}_k(t)}{\mathcal{E}_p(t)} = \frac{v^2(t)}{x^2(t)}.
\end{equation}
This property of the system is closely connected with another phase space property, namely with the phase angle
\begin{equation}
\phi = \arctan \frac{v}{x}
\end{equation}
The distribution $p(r)$ has
a universal asymptotics, and this universality is closely connected with non-independence of the velocity and coordinate processes, see Eq.~(\ref{JCF}) and \cite{sokolov2010}.

Imagine the distribution of the phase angle is known and is given by a PDF $p(\phi)$. Then $r = \tan^2 \phi$ and
\begin{eqnarray}
p(r) & = & p_\phi(\pm \phi(r))\left|\frac{d \phi}{dr} \right| \\ \nonumber
& = & \left[p_\phi(\arctan\sqrt{r})+ p_\phi(-\arctan\sqrt{r})\right]\frac{1}{2(1+r)\sqrt{r}},
\end{eqnarray}
note that there are two solutions for $\phi$ for a given $r$.
For $r \to \infty$, the argument of $p_\phi$, $\arctan\sqrt{r}$, tends to $\pi/2$ and therefore the behavior of $p(r)$
depends on whether $p_\phi(\phi)$ does or does not have a singularity at $\phi = \pm \pi/2$.

In the case of a driven harmonic oscillator, the association between the velocity and the coordinate processes makes the
distribution $p_\phi(\phi)$ non-singular at $\pm \pi/2$, which can be seen from the expressions of the corresponding spectral measures
as given in \cite{zozor2011spectral} which do not show singularities at $\theta = \pm \pi/2$,
and can be grasped from the graphical representation of the corresponding level curves.
Therefore, for $r \to \infty$
\begin{equation}
p(r) \propto \frac{C}{(1+r)\sqrt{r}} \simeq \frac{1}{r^{3/2}}
\label{eq:energyratio}
\end{equation}
with $C = [p_\phi(-\pi/2)+p_\phi(\pi/2)]/2$,
and is independent of $\alpha$. Interestingly enough, the inverse of $r$, the ratio of the potential and the kinetic energy, has
exactly the same asymptotic distribution, as it is evident by the explicit change of variables.

The universal $r^{-3/2}$ asymptotics of $p(r)$, see Eq.~(\ref{eq:energyratio}), originates due to lack of independence between position and velocity, see Eq.~(\ref{JCF}).
If $x$ and $v$ were independent the behavior would be very different, due to the fact that the corresponding spectral
measure is concentrated (i.e. has singularities) at $\theta = 0, \pm \pi/2,\mbox{ and } \pi$, i.e. at the intersections of the unit sphere with the axes \cite{samorodnitsky1994}.
The distribution of $r$ can be derived from the distribution of a quotient of two
independent symmetric L\'evy-stable variables which possesses quite a complicated form \cite{rathie2016exact}.
The ratio $r$ of instantaneous kinetic $\mathcal{E}_k(t)$ and potential $\mathcal{E}_p(t)$ energies has non-universal asymptotics
\begin{equation}
 p(r) \simeq
 \left\{
 \begin{array}{lcl}
 \frac{1}{r^{1+\alpha/2}} & \mbox{for} & 0<\alpha<1 \\
 \\
 \frac{\ln r}{r^{\frac{3}{2}}} & \mbox{for} & \alpha=1 \\
 \\
 \frac{1}{r^{3/2}} & \mbox{for} & 1 < \alpha \leqslant 2 \\
 \end{array}
 \right.
 \label{eq:energy-independent},
\end{equation}
see Appendix~\ref{sec:app-ratio-energy}.
Eq.~(\ref{eq:energy-independent}) should be contrasted with the universal and correct asymptotics given by Eq.~(\ref{eq:energyratio}).
For $\alpha=2$, in the stationary state, position and velocity are independent. Therefore for $\alpha=2$ asymptotics predicted by Eq.~(\ref{eq:energyratio}) and~(\ref{eq:energy-independent}) are the same.

Figure~\ref{fig:ratio} presents the ratio $r$ of instantaneous kinetic $\mathcal{E}_k$ and potential $\mathcal{E}_p$ energy for $\gamma=1,2,10$ (from top to bottom). Various panels correspond to different values of the stability index $\alpha$.
Solid lines present theoretical asymptotic given by Eq.~(\ref{eq:energyratio}).

\section{Summary and conclusions \label{sec:summary}}

In the present work we extend earlier studies \cite{sokolov2010} of the damped harmonic oscillator driven by L\'evy noises  in phase space characterized by the position $x$ and velocity $v=\dot{x}$.
The stationary state is given by a 2-dim $\alpha$-stable density \cite{samorodnitsky1994}.
First of all, in stationary states, position and velocity are no longer independent \cite{sokolov2010}, leading to a considerable difference from the usual case of the Gaussian white noise.
The studied system, i.e. damped harmonic oscillator driven by $\alpha$-stable noise is a highly non-equilibrium system displaying unexpected properties. Presence of non-equilibrium external noise introduces dependence between position and velocity in the stationary state. This dependence is responsible for violation of basic concepts of equilibrium statistical mechanics.

Kinetic and potential energies of a harmonic oscillator driven by a symmetric $\alpha$-stable noise have the same power-law asymptotics of $\mathcal{E}^{-(1+\alpha/2)}$ type determined by the noise type.
Contrary to  the classical Gaussian case, showing the
equipartition between the kinetic and the potential energy, $\langle \mathcal{E}_k \rangle = \langle \mathcal{E}_p \rangle$, we demonstrate that no such equipartition in a whatever statistical sense is observed for the L\'evy noise, except for the case of vanishing damping. For small friction there is a ``kind of'' stochastic equipartition, i.e. the potential and the kinetic energy distributions in the limit of $\gamma \to 0$ have the same widths. With the increasing damping larger fraction of energy is stored in the form of the potential energy.
Consequently, with increasing $\gamma$ the ratio of kinetic and potential energy distributions' widths decreases, and is given by a power-law in $\gamma$ with the exponent depending on the stability index $\alpha$.
In the limit of $\gamma \to \infty$ the system is fully overdamped. In such a case stochastic oscillator is fully characterized by its position only \cite{chechkin2002}, and the kinetic energy vanishes.

Finally, we have studied  the distribution of the ratio $r= \mathcal{E}_k(t)/\mathcal{E}_p(t) $ of instantaneous kinetic and potential energies in the stationary state. We show that this ratio has a universal $r^{-3/2}$ asymptotics independent on the stability index $\alpha$, which
differs strikingly from the situation when the position and the velocity of the oscillator were independent.


\begin{acknowledgments}
 This project has been supported in part by grants from National Science Center (2014/13/B/ST2/02014) and from the Institute of Physics (Focus 123/F/BD/2016).
Computer simulations have been performed at the Academic
Computer Center Cyfronet, Akademia G\'orniczo-Hutnicza (Krak\'ow, Poland).
\end{acknowledgments}

\appendix

\section{Units \label{sec:app-units}}

The white L\'evy noise $\zeta_\alpha(t)$ is by definition a formal derivative of the strictly $\alpha$-stable  L\'evy motion process $\zeta_\alpha(t)\equiv \frac{dL_{\alpha}(t)}{dt}$ whose increments are independent stationary variables distributed according to the symmetric $\alpha$-stable density, see Eq.~(\ref{eq:char}). The self-similarity of the L\'evy motion (L\'evy process) signifies that
its realizations fulfill the condition $L_{\alpha}(t)=t^{1/\alpha} L_\alpha(1)$. In view of the above, scaling properties of the L\'evy white noise assume the change of variables according to $\zeta_{\alpha}(t_0t)\rightarrow t_0^{1/\alpha-1} \zeta_\alpha(t)$.

\section{Ratio of distribution widths $D$ \label{sec:app-ratio-width}}

Small ($\gamma \to 0$) and large ($\gamma \to \infty$) asymptotics of the quotient $D=\frac{w_v^\alpha}{w_x^\alpha}$, see Eq.~(\ref{eq:quotient}), can be calculated analytically.
For $\gamma \to 0$ the system is strongly underdamped, for which
\begin{eqnarray}
G(t) & = & \frac{\exp(-\gamma t/2)}{\sqrt{1-\gamma^2/4}}\sin\left[ \sqrt{1-\gamma^2/4}t \right] \\ \nonumber
& \simeq & \exp(-\gamma t/2) \sin t,
\end{eqnarray}
and
\begin{equation}
G_v(t) \simeq \exp(-\gamma t/2) \left[-\frac{ \gamma}{2} \sin t + \cos t\right]
\end{equation}
in the lowest order in $\gamma$. The last expression can be rewritten as
\begin{equation}
G_v(t) \simeq \exp(-\gamma t/2) \sqrt{1 + \frac{\gamma^2}{4}} \cos(1+\phi)
\end{equation}
with $\phi = \arccos(1/\sqrt{1+{\gamma^2}/{4}})$.

For $\gamma \to 0$ we have
\begin{eqnarray}
w_x^\alpha & \simeq & \sigma^\alpha \int_0^\infty | \exp(-\gamma t/2) \sin t|^\alpha dt \\ \nonumber
& = & \sigma^\alpha \int_0^\infty \exp(-\gamma \alpha t/2)
|\sin t|^\alpha dt,
\end{eqnarray}
and a similar expression (with a cosine) for $w_v^\alpha$. For $\gamma \to 0$ the exponential hardly changes on the period of oscillations of the trigonometric function, so we can average over these oscillations. Essentially what we do is to split the domain of integration
into the $\pi$-intervals, which are the domains of periodicity of the absolute value of the trigonometric function and
rewrite the total integral as the sum
\begin{eqnarray}
w_x^\alpha &=& \sum_{n=0}^\infty \int_{n\pi}^{(n+1) \pi} \exp(-\gamma \alpha t/2) |\sin t|^\alpha dt \\ \nonumber
&=& \sum_{n=0}^\infty e^{n\pi \gamma \alpha /2} \int_{0}^{\pi} \exp(-\gamma \alpha t/2) |\sin t|^\alpha dt \\ \nonumber
&=& \frac{\exp(-\gamma \alpha t^*/2)}{1-\exp(\pi \gamma \alpha /2)} \int_{0}^{\pi} |\sin t|^\alpha dt
\end{eqnarray}
with $0 < t^* < \pi$.
The expression for $w_v$ in the lowest order in $\gamma$ is
\begin{equation}
w_v^\alpha = \frac{\exp(-\gamma \alpha t^{**}/2)}{1-\exp(\pi \gamma \alpha /2)} \int_{0}^{\pi} |\cos(t+\phi)|^\alpha dt,
\end{equation}
and only differs with respect to the position of the intermediate point $0< t^{**} < \pi$. The integrals over the trigonometric functions are the same, and are given by
\begin{equation}
\int_{0}^{\pi} |\sin t|^\alpha dt = \int_{0}^{\pi} |\cos(t+\phi)|^\alpha dt = \sqrt{\pi} \frac{\Gamma[(\alpha+1)/2]}{\Gamma(\alpha/2 +1)}.
\end{equation}
Therefore
\begin{equation}
D=\frac{w_v^\alpha}{w_x^\alpha} = \exp[-\gamma \alpha (t^{**}-t^*)/2] \to 1
\end{equation}
for $\gamma \to 0$. For small friction we have a ``kind of'' equipartition, i.e. both densities $p(x)$ and $p(v)$ are characterized by the same width. In the lowest order in $\gamma$ this can be obtained by expanding the denominator:
\begin{equation}
w_v^\alpha = w_x^\alpha = \frac{2}{\sqrt{\pi}\gamma \alpha} \frac{\Gamma[(\alpha+1)/2]}{\Gamma(\alpha/2 +1)}.
\end{equation}
For $\gamma \to \infty$ we start from the explicit expression
\begin{eqnarray}
\label{eq:greenas}
G(t) &=& \frac{\exp(-\gamma t/2)}{\sqrt{\gamma^2/4-1}}\sinh \left[\sqrt{\gamma^2/4-1}\;t \right] \\ \nonumber
&=& \frac{1}{2\sqrt{\gamma^2/4-1}}\left[e^{(\sqrt{\gamma^2/4-1}-\gamma/2) t} - e^{-(\sqrt{\gamma^2/4-1}+\gamma/2) t} \right] \\ \nonumber
&\simeq& \frac{1}{\gamma} \left(e^{-\frac{t}{\gamma}} - e^{-\gamma t} \right)
\end{eqnarray}
where in the last line only the leading contributions in $\gamma$ in the exponential are retained.
For $\gamma \to \infty$ the first exponential is decaying very slowly, and gives the major contribution to the time integral
\begin{equation}
G(t) \simeq \frac{1}{\gamma}e^{-\frac{t}{\gamma}}.
\end{equation}
We now can estimate the integral for $w_x^\alpha$, and get
\begin{equation}
w_x^\alpha \simeq \frac{\gamma^{1-\alpha}}{\alpha}.
\end{equation}
Similarly for $G_v(t)= \frac{d}{dt}G(t)$ we get
\begin{equation}
G_v(t) \simeq \frac{1}{\gamma^2}e^{-\frac{t}{\gamma}} + e^{-\gamma t}.
\label{eq:gv}
\end{equation}
In order to calculate $w_v^\alpha$ special care is required.
The second term in Eq.~(\ref{eq:gv}) cannot be neglected because for $\gamma \gg 1$ it is larger than the first one
\begin{equation}
 w_v^\alpha = \int_0^\infty |G_v(t)|^\alpha dt = \int_0^\infty \left[ \frac{1}{\gamma^2}e^{-\frac{t}{\gamma}} + e^{-\gamma t} \right]^\alpha dt.
 \label{eq:integrand}
\end{equation}
The integrand (\ref{eq:integrand}) shows a crossover between two types of asymptotics: for $\gamma \gg 1$ the short time behavior is dominated by the fast decay $e^{-\alpha \gamma t}$, while at long times it is dominated by the slow decay $\gamma^{-2}e^{-\alpha t/\gamma}$.
The crossover between two regimes takes place at time $t_c=2\gamma^{-1}\ln\gamma$. Consequently, the total integral can be estimated as
\begin{equation}
 w_v^\alpha \simeq \int_0^{t_c} e^{-\alpha \gamma t} dt + \gamma^{-2\alpha}\int_{t_c}^\infty e^{-\frac{\alpha t}{\gamma} } dt
\end{equation}
resulting in
\begin{equation}
 w_v^\alpha \simeq \frac{1}{\alpha \gamma} \left[ 1 - \frac{1}{\gamma^{2\alpha}} \right] +\frac{\gamma^{1-2\alpha}}{\alpha}\gamma^{2\alpha/\gamma^2}.
\end{equation}
For $\gamma \gg 1$ the above expression can be further simplified to
\begin{equation}
 w_v^\alpha \simeq \frac{1}{\alpha \gamma} +\frac{\gamma^{1-2\alpha}}{\alpha}
\end{equation}
leading to the dominating terms
\begin{equation}
w_v^\alpha \simeq
\left\{
\begin{array}{clc}
 \frac{\gamma^{1-2\alpha}}{\alpha} & \mbox{for} & 0 < \alpha < 1 \\
\frac{1}{\alpha\gamma} & \mbox{for} & 1 \leqslant \alpha \leqslant 2 \\
\end{array}
\right..
\end{equation}
Finally, the ratio of distributions widths scales as
\begin{equation}
 D= \frac{w_v^\alpha}{w_x^\alpha}=
 \left\{
 \begin{array}{lcl}
 \gamma^{-\alpha} & \mbox{for} & 0< \alpha <1 \\
 \gamma^{\alpha-2} & \mbox{for} & 1 \leqslant \alpha \leqslant 2 \\
 \end{array}
 \right..
 \label{eq:dscalinga}
\end{equation}
For $\alpha=2$, from Eq.~(\ref{eq:greenas}) and the definition
\begin{equation}
 D=1,
\end{equation}
as predicted by Eq.~(\ref{eq:dscalinga}).

\section{Ratio of instantaneous kinetic and potential energies  $r$ \label{sec:app-ratio-energy}}

For pedagogical reason it is interesting to assume that $v$ and $x$ are independent. In such a case, the qualitative discussion of the asymptotic behavior of $r=\mathcal{E}_k(t)/\mathcal{E}_p(t)$ is however very simple.
The large values of
\begin{equation}
z = \frac{x}{y}
\end{equation}
will typically appear either due to the very large values of the enumerator or to very small values of denominator
(the possibility that both occurs simultaneously is very small, and plays the role only in the case of the Cauchy distribution,
\textit{vide infra}). In the symmetric L\'evy case the probability density of large values of numerator decays as
\begin{equation}
p(x) \simeq \frac{1}{|x|^{1+\alpha}},
\end{equation}
while the distribution of $q=1/y$ is given by the variable transformation
\begin{equation}
p(q)= p\left(\frac{1}{q}\right)\frac{1}{q^2}.
\end{equation}
Since $p(x)$ is non-singular and does not vanish at zero, the tail of the PDF $p(q)$ is
universal and of the same type as for the Cauchy distribution:
\begin{equation}
p(q) \propto \frac{1}{q^2}.
\end{equation}
For $\alpha >1$ the tail of $p(z)$ is dominated by the tail of $p(q)$ and therefore $p(z) \propto z^{-2}$.
The variable transformation to $r = z^2/2$ transforms this tail into
\begin{equation}
p(r) = \frac{1}{r^{3/2}},
\end{equation}
exactly as in the case of the correlated variables above. An explicitly solvable example is given by the Gaussian case
$\alpha = 2$ for which the distribution of $z$ is known explicitly: it is a Cauchy distribution
\begin{equation}
p(z) = \frac{1}{\pi} \frac{1}{1+z^2}.
\end{equation}
The variable transformation to $r$ gives
\begin{equation}
p(r) = \frac{1}{\pi}\frac{1}{(r+1)\sqrt{r}}.
\end{equation}

For $\alpha < 1$ the tail of the ratio $z$ is dominated by the tail of the enumerator, so that
\begin{equation}
p(z) \simeq \frac{1}{|z|^{1+\alpha}},
\end{equation}
and the variable transformation gives
\begin{equation}
p(r) \simeq \frac{1}{r^{1+\frac{\alpha}{2}}}.
\end{equation}
The transition between the two regimes happens at $\alpha = 1$, i.e. for the Cauchy distribution, for which the distribution
of the ratio of the two variables is again explicitly known \cite{ride1965distributions}
\begin{equation}
p(z)=\frac{1}{\pi^2} \frac{1}{z^2-1} \ln z^2,
\end{equation}
so that
\begin{equation}
p(r) =\frac{1}{2 \pi^2} \frac{1}{(r-1)\sqrt{r}} \ln r
\end{equation}
and involves a logarithmic correction: its asymptotic behavior is
\begin{equation}
p(r) \simeq \frac{\ln r}{r^{\frac{3}{2}}}.
\end{equation}
Consequently, if the velocity $v$ and position $x$ would be independent, the ratio of instantaneous kinetic $\mathcal{E}_k$ and potential $\mathcal{E}_p$ energy, has the following non-universal asymptotics
\begin{equation}
 p(r) \simeq
 \left\{
 \begin{array}{lcl}
 \frac{1}{r^{1+\alpha/2}} & \mbox{for} & 0<\alpha<1 \\
 \\
 \frac{\ln r}{r^{\frac{3}{2}}} & \mbox{for} & \alpha=1 \\
 \\
 \frac{1}{r^{3/2}} & \mbox{for} & 1 < \alpha \leqslant 2 \\
 \end{array}
 \right..
\end{equation}
Contrary to this $\alpha$-dependent behavior, the correct asymptotic behavior of $p(r)$ for the harmonic L\'evy oscillator, where $v$ and $x$ are not independent, is universal:
\begin{equation}
 p(r) \simeq r^{-3/2}.
\end{equation}

%
%

\def\url#1{}\def\url#1{}

\end{document}